\documentclass[aps,prd,twocolumn,superscriptaddress,nofootinbib,10pt]{revtex4-1}
\usepackage[paperwidth=21cm,paperheight=29.7cm,top=2.54cm,bottom=2.54cm,left=2cm,right=2cm]{geometry}
\usepackage[colorlinks,linkcolor=blue,anchorcolor=blue,citecolor=blue,urlcolor=blue]{hyperref}
\usepackage{graphicx,subfigure}%dcolumn
\usepackage{xcolor}% Explicit support for revision colors.
\usepackage[figuresright]{rotating}
\usepackage{amsmath,amsfonts,amssymb,bm}
\usepackage{array,enumitem,multirow}%,makecell
\usepackage{acronym}
\usepackage{makecell}
\newcommand{\be}{\begin{equation}}
\newcommand{\ee}{\end{equation}}
\newcommand{\bea}{\begin{eqnarray}}
\newcommand{\eea}{\end{eqnarray}}
\newcommand{\dd}{\mathrm{d}}% Upright differential with revision marking.

\def\ccr#1{{\color{red} #1}}
\def\ccb#1{{\color{blue} #1}}

\newacro{GR}{general relativity}
\newacro{GW}{gravitational wave}
\newacro{MG}{modified gravity theory}
\newacro{BH}{Black hole}
\newacro{PN}{post-Newtonian}
\newacro{ppE}{parameterized post-Einsteinian}
\newacro{GCB}{galactic ultra-compact binary}
\newacro{SBHB}{stellar-mass black hole binary}
\newacro{MBHB}{massive black hole binary}
\newacro{BHB}{black hole binary}
\newacro{IMBHB}{intermediate-mass black hole binary}
\newacro{EMRI}{extreme mass ratio inspiral}
\newacro{IMRI}{intermediate mass ratio inspiral}
\newacro{SGWB}{stochastic gravitational wave background}
\newacro{MECO}{minimal energy circular orbit}
\newacro{FAR}{false alarm rate}
\newacro{CE}{Cosmic Explorer}
\newacro{ET}{Einstein Telescope}
\newacro{LISA}{Laser Interferometer Space Antenna}
\newacro{EdGB}{Einstein-dilaton Gauss-Bonnet}
\newacro{dCS}{dynamic Chern-Simons}
\newacro{SNR}{signal-to-noise ratio}
\newacro{FIM}{Fisher Information Matrix}
\newacro{ISCO}{innermost stable circular orbit}
\newacro{NSBH}{neutron star-black hole binary}
\newacro{MCMC}{Markov Chain Monte Carlo}
\newacro{QNM}{quasinormal mode}
\begin{document}

\title{\texorpdfstring{Negative Dynamical Friction in Dark Matter Spikes around Rotating Black Holes}{Negative Dynamical Friction in Dark Matter Spikes around Rotating Black Holes}}% Force line breaks with \\

\author{Wei Xiong}
\email{202210187053@mail.scut.edu.cn}%
  
\author{Peng-Cheng Li}%
\thanks{\href{mailto:pchli2021@scut.edu.cn}{pchli2021@scut.edu.cn}, corresponding author}
\affiliation{%
	School of Physics and Optoelectronics, South China University of Technology, Guangzhou 510641, China.
}%
 \author{Bin Chen}
 \thanks{\href{mailto:chenbin1@nbu.edu.cn}{chenbin1@nbu.edu.cn}, corresponding author}
\affiliation{Institute of Fundamental Physics and Quantum Technology, and School of Physical Science and Technology, Ningbo University, Ningbo, Zhejiang 315211, China}
\affiliation{Zhejiang Key Laboratory of Extreme Universe, and BINGO Center,
	Ningbo University, Ningbo, Zhejiang 315211, China}
\affiliation{
	School of Physics, and Center for High Energy Physics, Peking University, No.~5 Yiheyuan Rd, Beijing 100871, China}
\date{\today}

\begin{abstract}
Dark matter may affect extreme-mass-ratio inspirals (EMRIs) through orbital-energy exchange and a conservative metric response. Here we show that dark matter spikes around rotating massive black holes can drive negative dynamical friction, supplying orbital energy to secondary black holes on stable prograde circular orbits.  The energy-gain region extends beyond the innermost stable circular orbit once the black-hole spin exceeds a model-dependent threshold. For an illustrative four-year inspiral with Milky Way--like halo parameters and mass ratio $q=10^{-5}$, the total dynamical-friction contribution produces a phase difference of order $10^2$ radians relative to vacuum. For the same system, the fixed-frequency environmental redshift correction is comparable in magnitude and opposite in sign to the first-order conservative self-force correction. Both environmental energy exchange and the conservative metric response are therefore essential to precision EMRI modeling.
\end{abstract}
%, both described here from a common relativistic phase-space distribution constructed through adiabatic black-hole growth.

\maketitle

%\tableofcontents

%\section{Introduction}

{\textit{Introduction}} ---
Extreme-mass-ratio inspirals (EMRIs) are key sources for future space-based gravitational-wave (GW) observatories~\cite{LISA:2022yao,TianQin:2015yph,Ruan:2018tsw}. Their long-lived signals can carry imprints of accretion disks~\cite{Abramowicz:2011xu,Speri:2022upm,Barausse:2007ph}, ultralight boson clouds~\cite{Brito:2015oca,Brito:2023clouds,Duque:2023seg,Dyson:2025dlj,Li:2026foy}, and dark matter (DM) halos or spikes~\cite{Gondolo:1999ef,Sadeghian:2013laa,Ferrer:2017xwm} around massive black holes (BHs)~\cite{Barausse:2014tra,LISA:2022kgy,Luo:2025ewp}. DM can modify the background geometry and bound orbits~\cite{Cardoso:2021wlq,Jusufi:2022jxu,Konoplya:2022hbl,Maeda:2024tsg,Shen:2023erj,Shen:2024qbb,Chakraborty:2024gcr,Yue:2026gju,Polcar:2022bwv,Dai:2023cft,Zhao:2024bpp,Yuan:2025qxz,Chen:2026myf,Zhao:2026yis}, drive dynamical friction and accretion~\cite{Yue:2017iwc,Li:2021pxf,Karydas:2024fcn,Zhou:2024vhk}, and alter GW generation and propagation~\cite{Cardoso:2022whc,Figueiredo:2023gas,Speeney:2024mas,Rahman:2025mip,Datta:2025}. These effects motivate efforts to infer DM properties from EMRI signals~\cite{Coogan:2021uqv,Zhang:2024ugv,Wang:2025rrj,Yuan:2026wmi,Li:2026hyn}. Interpreting these imprints requires environmental corrections to be assessed at the accuracy demanded by precision EMRI waveforms~\cite{LISAConsortiumWaveformWorkingGroup:2023arg}.

For cold particle DM, progress toward this goal includes modeling the companion-induced evolution of DM spikes~\cite{Kavanagh:2020cfn,Kavanagh:2024lgq} and incorporating relativistic corrections to dynamical friction~\cite{Speeney:2022ryg,Mitra:2025tag}. Complementary approaches describe environmental effects on gravitational radiation through coupled matter--metric perturbations~\cite{Cardoso:2022whc,Datta:2025}. For nonrotating BHs, Rahman and Takahashi~\cite{Rahman:2025mip} developed a post-adiabatic framework incorporating conservative orbital corrections and modified GW fluxes, while treating dynamical friction with an approximate prescription.

A fully relativistic framework for computing dynamical friction in collisionless environments has been developed for Kerr backgrounds, with explicit applications to DM spikes around Schwarzschild BHs~\cite{Vicente:2025gsg}. Regarding conservative effects, rotating BH--matter geometries have been constructed~\cite{Fernandes:2025osu}, and the stress-energy tensor and metric response of Schwarzschild DM spikes have been calculated from the particle phase-space distribution~\cite{Xiong:2025zqp}. Rotating environments can transfer orbital energy to an embedded perturber, as illustrated by studies of stellar disks~\cite{Bonetti:2020}. For Kerr DM spikes, the sign and strength of this exchange depend on the relativistic distribution of bound particle orbits. The same distribution also sources a conservative metric response, whose significance for precision EMRI modeling can be assessed against the vacuum first-order conservative self-force~\cite{LISAConsortiumWaveformWorkingGroup:2023arg}.

In this Letter, we show that negative dynamical friction can arise in a DM spike, leading to the supply of orbital energy to a secondary BH on prograde equatorial circular orbits outside the innermost stable circular orbit (ISCO) once the BH spin exceeds a model-dependent threshold. Particles overtaking the companion in azimuth can supply orbital energy through scattering. The asymmetry of allowed Kerr orbits, combined with enhanced scattering at small relative speeds, can make this supply exceed the losses \ccb{due} to slower particles. We further show that this energy gain does not require a finely tuned phase-space distribution and occurs for spikes grown from both cusped and cored halos. We combine the kinetic approach of Ref.~\cite{Xiong:2025zqp}  with the relativistic scattering framework of Ref.~\cite{Vicente:2025gsg} to obtain dynamical friction and the stationary metric response from a common distribution. The distribution follows from adiabatic BH growth~\cite{Ferrer:2017xwm}  and includes generic bound particle orbits. The resulting stress-energy tensor retains particle kinetic energy and radial pressure and satisfies the standard energy conditions because the distribution is nonnegative. For Milky Way--like halo parameters and $q=10^{-5}$, we quantify the accumulated dynamical-friction phase shift and show that the fixed-frequency environmental redshift correction is comparable to the first-order conservative self-force correction~\cite{Kavanagh:2016Redshift}.

\textit{Kinetic framework and dark matter spike}---
We work in units with $G=c=1$ and consider a BH of mass $M$, a collisionless DM spike, and a secondary BH of mass $m$. A distribution of DM particles on generic bound Kerr geodesics determines both the stress-energy sourcing the stationary metric response and the local momentum transfer to the secondary. This description retains particle kinetic energy and radial pressure. Our inspiral calculations specialize to equatorial quasi-circular secondary orbits with four-velocity $\bm u$ and proper time $\tau$.

We expand the full metric $\bm g$ and invariant number distribution $\bm{\mathcal F}$ in independent parameters $\epsilon$ for the environment and $q=m/M\ll1$ for the secondary, without imposing a hierarchy between them. The expansion assumes only that the environmental metric response is weak in the modeled domain~\cite{Xiong:2025zqp}; it does not require the total halo mass to be smaller than $M$. We write
\begin{eqnarray}
	\bm{g}_{\mu\nu} &=& g_{\mu\nu} + \epsilon \ h^{(1,0)}_{\mu\nu} + q \ h^{(0,1)}_{\mu\nu} + \epsilon \ q \ h^{(1,1)}_{\mu\nu} + \dots \\
	\bm{\mathcal{F}} &=& \epsilon \ \mathcal{F}^{(1,0)} + \epsilon \ q \ \mathcal{F}^{(1,1)} +  \epsilon^2  \mathcal{F}^{(2,0)} + \ldots. 
	\label{eq:environment_perturbations}
	%\bm{u} &=& \bar{u}^{\mu} + \epsilon \ u^{\mu}_{(1,0)} + q \ u^{\mu}_{(0,1)} + \ldots .
\end{eqnarray}
The zeroth-order metric $g_{\mu\nu}$ is Kerr, and the secondary follows a timelike geodesic. The linearized Einstein equations are
\begin{eqnarray}
	G^{(1,0)}_{\mu\nu} [h^{(1,0)}_{\mu\nu}] &-&  8 \pi \ \mathcal{T}^{(1,0)}_{\mu\nu}[\mathcal{F}^{(1,0)}] =0 \label{eq:Einstein_10_environment} \\
	G^{(0,1)}_{\mu\nu} [h^{(0,1)}_{\mu\nu}] &-&  8 \pi \ T^{(0,1)}_{\mu\nu}[\bm{u}] =0, \label{eq:Einstein_01_companion}
\end{eqnarray}
where $\mathcal T_{\mu\nu}^{(1,0)}$ and $T_{\mu\nu}^{(0,1)}$ are the leading environmental and secondary stress-energy tensors, respectively.

The DM particles have rest mass $\mu$, four-velocity $\bm v$, and momentum $\bm p=\mu\bm v$. The leading number distribution $\mathcal F^{(1,0)}$ yields the stress-energy tensor~\cite{Fackerell:1968}
\begin{equation}
	\mathcal{T}^{(1,0)}_{\mu\nu} = \int_{\mathcal{B}(x)} p_{\mu} p_{\nu} \  \mathcal{F}^{(1,0)} \sqrt{-g} \ \dd ^4p.
	\label{eq:integral_energy_momentum}
\end{equation}
Here $\mathcal B(x)$ denotes the momentum-space domain of bound, future-directed timelike Kerr geodesics through $x$; $\mathcal F^{(1,0)}$ includes the fixed-mass delta function specified in the End Matter. The stationary distribution $\mathcal F^{(1,0)}$ depends only on the particles' conserved quantities. It therefore satisfies the collisionless Boltzmann equation and ensures $\nabla^\mu\mathcal T_{\mu\nu}^{(1,0)}=0$ on the Kerr background.

\begin{figure}[htbp]
	\centering
	\includegraphics[width = 0.482\textwidth]{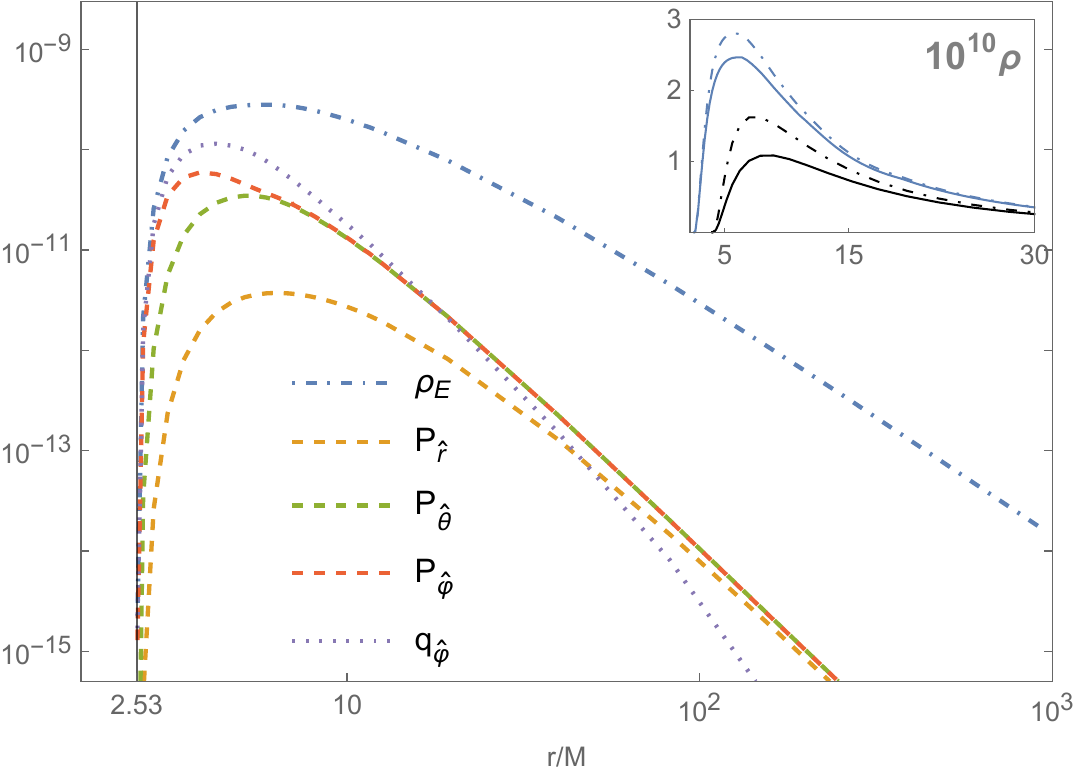}
	\\[0.3cm]
	\includegraphics[width = 0.48\textwidth]{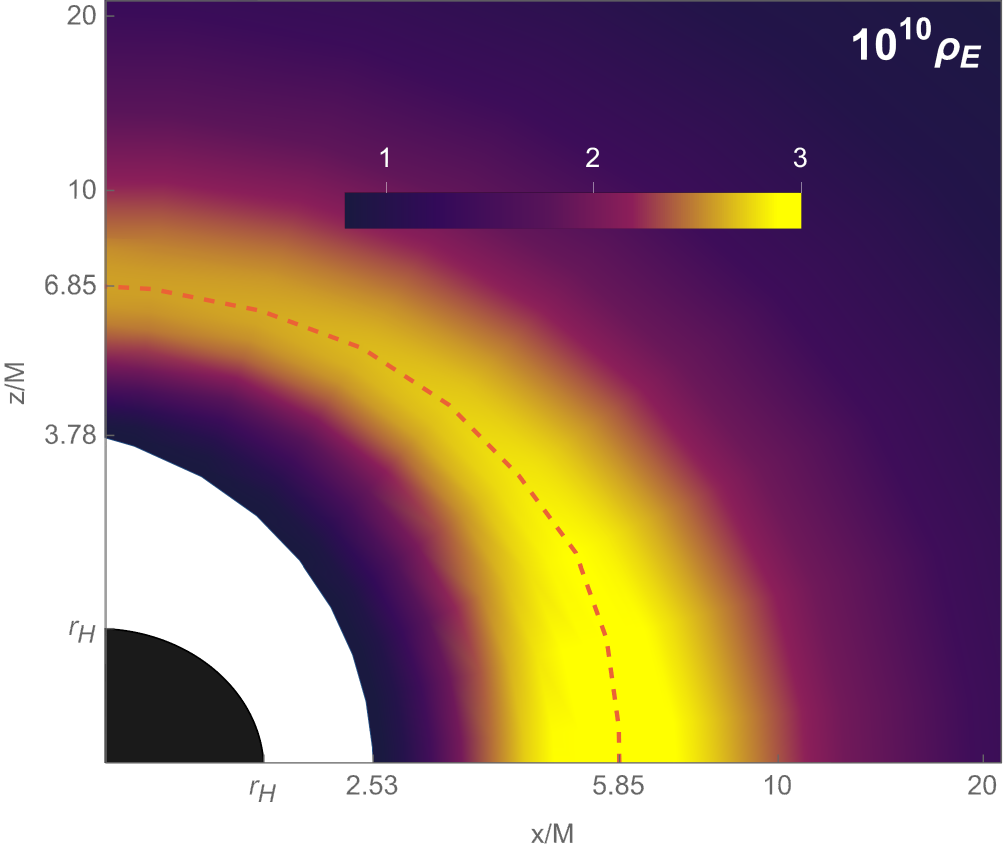}
	\caption{Properties of the DM spike for the parameters in Eq.~(\ref{eq:Milky Way-like}). Upper: equatorial profiles of energy density $\rho_E$, pressures $P_{\hat r},P_{\hat\theta},P_{\hat\varphi}$, and azimuthal energy flux $q_{\hat\varphi}$, as measured by a zero angular momentum observer (ZAMO)~\cite{Bardeen:1972}. All vanish at the prograde marginally bound orbit (MBO), the spike's inner edge $r_{\mathrm{MBO}}^{(+)}$. Inset: energy density (dot-dashed) and invariant rest-mass density (solid; defined in the End Matter) for rotating (blue) and nonrotating (black) BHs. Lower: energy density in the $(x,z)=(r\sin\theta,r\cos\theta)$ plane. Black and white indicate the BH interior and the vacuum region inside the spike edge, respectively; the red dashed curve fits the radial density maxima.}
	\label{figProfile}
\end{figure}

The kinetic integrals are performed on the Kerr background, whose covariant derivative is denoted by $\nabla$. In Boyer--Lindquist coordinates $(t,r,\theta,\varphi)$, the companion's specific orbital energy, Carter constant, and axial angular momentum are
\begin{equation}
	E \equiv - \bm{u} \ \cdot \ \bm{\partial}_t, \ \ \ C \equiv \bm{K}(\bm{u},\bm{u}), \ \ \ L_z \equiv  \bm{u} \ \cdot \ \bm{\partial}_\varphi,
	\label{eq:conserved_quantities_companion}
\end{equation}
where $K$ is the Kerr Killing tensor \cite{Walker:1970un}. Replacing $u^\mu$ with $v^\mu$ defines the particle quantities $(\mathcal E,\mathcal{C},l_z)$. We integrate Eq.~(\ref{eq:integral_energy_momentum}) over these quantities; the mass normalization and mass-shell Jacobian are given in the End Matter.

We construct the spike by adiabatically growing the central BH in a Hernquist halo, $\rho_M^{\textrm{(H)}}=\frac{M_0}{2\pi}\frac{r_0}{r(r+r_0)^3}$~\cite{Hernquist:1990be}, with halo mass $M_0$ and scale radius $r_0$. Its reduced number distribution $f^{(H)}(\hat{\mathcal E})$, with $\hat{\mathcal E}$ the Newtonian binding energy per unit mass, follows from Abel inversion~\cite{Binney:2011}.  Conservation of orbital actions maps this distribution to the final spike distribution $f^{(1,0)}$~\cite{Ferrer:2017xwm,Gondolo:1999ef,Sadeghian:2013laa}, as detailed in the End Matter. The post-growth actions are evaluated in the Kerr geometry, so the construction is intended for the BH-dominated region. We adopt Milky Way--like halo parameters
\begin{eqnarray}
    &&r_{0} = 227775 \ M_0, \ \ M_{0} = 387500 \ M, \ \ \nonumber \\
	&&a = 0.65M , \ \ M=4\times10^6 M_{\odot}.
    \label{eq:Milky Way-like}
\end{eqnarray}
following Refs.~\cite{Eadie:2015,Dokuchaev:2013xda}, and fix $q=10^{-5}$. The physical BH mass sets the time scale of the inspiral. We evaluate Eq.~(\ref{eq:integral_energy_momentum}) by Monte Carlo integration over bound, noncaptured Kerr geodesics~\cite{Ferrer:2017xwm}.

Figure~\ref{figProfile} shows the ZAMO-measured stress-energy components for the parameters in Eq.~(\ref{eq:Milky Way-like}). The nonnegative distribution of massive particles guarantees that the leading kinetic tensor satisfies the null, weak, strong, and dominant energy conditions (see End Matter). A Hawking--Ellis eigenvalue analysis~\cite{Hawking:1973uf,Kolassis:1988} confirms its type-I structure and provides a numerical consistency check. The energy density dominates all other components. As in the nonrotating case~\cite{Xiong:2025zqp}, $P_{\hat r}$ is nonzero: near the density maximum it is more than an order of magnitude below the other nonzero components, but at $r/M\sim10-100$ it becomes comparable to the tangential pressures. The spin-induced azimuthal energy flux, absent in the static limit, exceeds these pressures in the strong-field region and decreases rapidly with radius. The inset shows that rotation enhances the density while reducing the relative excess $(\rho_E-\rho_M)/\rho_M$ over the invariant rest-mass density.

The lower panel shows that the density is concentrated toward the equatorial plane and suppressed toward the rotation axis. Axisymmetry and equatorial reflection symmetry allow us to display only $x,z\geq0$.

We solve Eq.~(\ref{eq:Einstein_10_environment}) with a stationary, axisymmetric metric ansatz satisfying circularity~\cite{Herdeiro:2015gia}, obtaining $h_{\mu\nu}^{(1,0)}$ by a spectral method~\cite{Fernandes:2022gde}. The ansatz, boundary conditions, and numerical procedure are described in the End Matter.

Schematically, the companion obeys
\begin{eqnarray}
	(\nabla_{\bm{u}}\bm{u})^\mu =&& \epsilon \ \mathcal{A}^{\mu}_{\textrm{met}}[h^{(1,0)}_{\mu\nu}]+q \ \mathcal{A}^{\mu}_{\textrm{1stGSF}}[h^{R(0,1)}_{\mu\nu}]  \nonumber \\ 
	&& + \epsilon \ q \ \mathcal{A}^{\mu}_{\textrm{DF}}[\mathcal{F}^{(1,0)},\bm{u}]  + \ldots,
	\label{eq:companion_motion}
\end{eqnarray}
where the three terms describe the conservative metric response, first-order vacuum gravitational self-force, and dynamical friction, respectively. Here $h^{R(0,1)}_{\mu\nu}$ is the regular field associated with the retarded perturbation, so the first-order self-force includes both conservative and dissipative contributions \cite{Pound:2012nt,Barack:2018yvs}. The ellipsis includes higher-order self-force terms starting at $O(q^2)$ and other mixed environmental contributions.

\textit{Negative dynamical friction}---
We evaluate momentum exchange in the companion's local rest frame using the Marck tetrad $(\bm u,\bm e_{\bar i})$~\cite{Marck:1983}. The transformation of particle velocities from Boyer--Lindquist coordinates to this frame is valid to generic companion orbits. In the companion's body zone, DM particles are treated as incident on an isolated BH, with capture and scattering cross sections $\sigma_{\textrm{acc}}(\bar v)$ and $\sigma_{\textrm{scat}}(\bar v)$ proportional to $m^2$. The particle speed is $\bar v=\sqrt{v^{\bar i}v_{\bar i}}$, and its four-velocity has tetrad components $\gamma_{\bar v}(1,v^{\bar i})$, with $\gamma_{\bar v}=(1-\bar v^2)^{-1/2}$. The corresponding acceleration is~\cite{Vicente:2025gsg}
\begin{eqnarray}
	\epsilon \ q \ \mathcal{A}^{\mu}_{\textrm{DF}} \triangleq && \frac{1}{m} \int_{\mathcal{B}(x)}  \left[\bar{v} \ \sigma_{\textrm{acc}} + \bar{v} \ \sigma_{\textrm{scat}} \right] \  p^{\mu}_{\bot} \nonumber \\ 
	&& \times  \ \epsilon \ f^{(1,0)} \gamma_{\bar{v}}^5 \ \dd ^3v^{\bar{i}}, \label{eq:acceleration_dynamical_friction}
\end{eqnarray}
where $f^{(1,0)}$ is the reduced number distribution and $p^\mu_\bot=(\delta^\mu{}_\nu+u^\mu u_\nu)p^\nu=\mu\gamma_{\bar v}v^{\bar i}e^\mu_{\bar i}$ is the momentum projected into the companion's rest space. We use the cross sections for a Schwarzschild companion from Ref.~\cite{Vicente:2025gsg} to focus on the rotating environment. Explicit kinetic integrals are given in the End Matter.

The acceleration $a^\mu_{\textrm{DF}}\equiv\epsilon q\mathcal A^\mu_{\textrm{DF}}$ gives the rate of change of the specific orbital energy
\begin{equation}
	\dot{E}_{\rm DF} = -(\bm{a}_{\textrm{DF}}  \cdot  \bm{\partial}_t)/u^t,
	\label{eq:conserved_quantities_dot_dynamical_friction}
\end{equation}
where the dot represents the derivative with respect to $t$.

We evaluate Eq.~(\ref{eq:acceleration_dynamical_friction}) by Monte Carlo integration in local three-velocity space, checking consistency with integration over conserved quantities in the far zone.

\begin{figure*}[t!]
	\begin{center}
		\mbox{%
			\includegraphics[width = 0.34\textwidth]{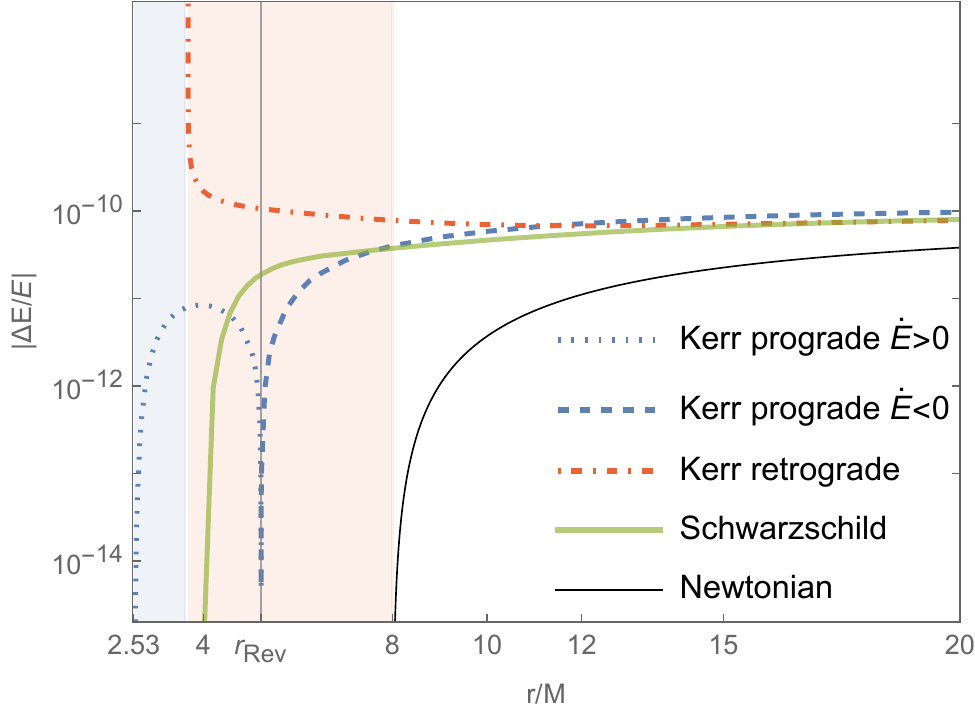}%
			\includegraphics[width = 0.33\textwidth]{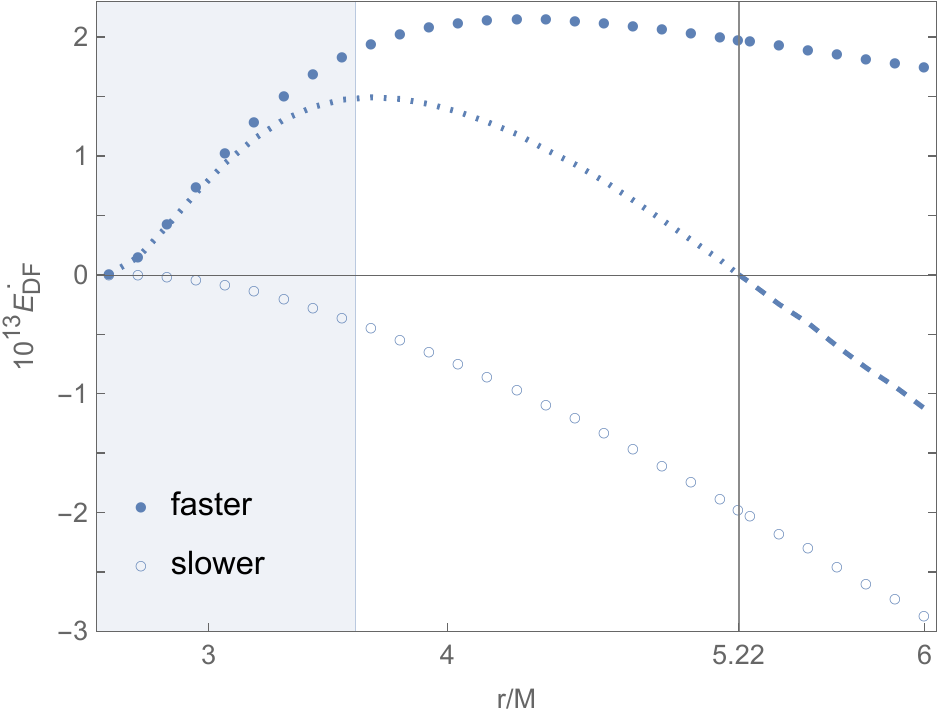}%
			\includegraphics[width = 0.33\textwidth]{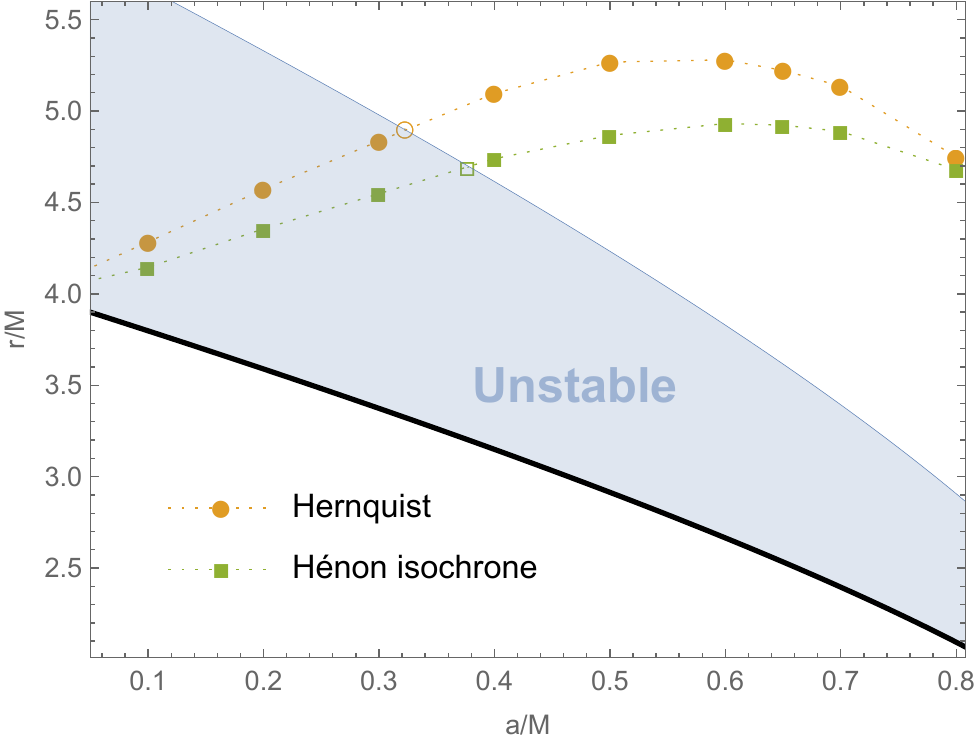}%
		}
		\vspace*{-0.5cm}
	\end{center}
	\caption{Energy transfer from environmental dynamical friction. Left: $|\Delta E_{\rm DF}/E|$ per circular orbit at $a/M=0.65$. Blue dotted/dashed segments indicate prograde gain/loss; retrograde Kerr, Schwarzschild~\cite{Vicente:2025gsg}, and Newtonian~\cite{Speeney:2022ryg} curves follow the legend. Blue/red shading marks unstable prograde/retrograde orbits. Middle: energy-rate contributions from locally faster ($v^{\bar3}>0$, filled) and slower ($v^{\bar3}<0$, open) particles; their net contribution (dotted/dashed) changes sign at $r_{\rm Rev}\simeq5.22M$ (vertical line). Right: $r_{\rm Rev}(a)$ for Hernquist (circles) and H\'enon isochrone (squares) spikes; the open circle marks the prograde ISCO crossing. Blue shading in the middle and right panels denotes unstable prograde orbits; the black curve is $r_{\mathrm{MBO}}^{(+)}(a)$.}
	\label{figDF}
\end{figure*}

For prograde circular orbits at $a/M=0.65$, Fig.~\ref{figDF} shows that dynamical friction supplies orbital energy between the prograde ISCO and $r_{\rm Rev}\simeq5.22M$. This energy gain arises when the contribution from locally faster particles outweighs the loss to slower particles.

To resolve these contributions, we split the integration domain into $\mathcal B_>=\mathcal B\cap\{v^{\bar3}>0\}$ and $\mathcal B_<=\mathcal B\cap\{v^{\bar3}<0\}$, where $v^{\bar3}=\bm v\cdot\bm e_{\bar3}/\gamma_{\bar v}$ is the local azimuthal three-velocity relative to the companion. Within our local scattering approximation, symmetry of the distribution under radial and polar velocity reversals makes the net momentum transfer purely azimuthal for equatorial circular companion orbits. The nonnegative weights in Eq.~(\ref{eq:acceleration_dynamical_friction}) imply $\dot E_>\geq0$ and $\dot E_<\leq0$. Their sum, $\dot E_{\rm DF}=\dot E_>+\dot E_<$, changes sign at $r_{\rm Rev}$ [Fig.~\ref{figDF} (middle)]. Kerr's binding and capture conditions make the allowed velocity domain asymmetric in the companion's frame. Enhanced scattering at small relative speeds can then make the forward force from overtaking particles exceed the drag from slower particles. The Supplemental Material demonstrates this mechanism with a constant phase-space distribution.

For comparison, Fig.~\ref{figDF} (left) plots the relative energy transfer per circular orbit, $\left|\Delta E_{\rm DF}/E\right|\equiv\left|(2\pi/\Omega)\dot E_{\rm DF}/E\right|$, for Kerr, Schwarzschild, and Newtonian calculations. The Schwarzschild and retrograde Kerr contributions remain dissipative for the cases shown. At larger radii, the relativistic results are comparable to each other and substantially exceed the Newtonian estimate~\cite{Speeney:2022ryg}. Results in the shaded unstable-orbit intervals diagnose the local interaction and do not represent adiabatic inspiral trajectories. The sharp increase in the retrograde energy transfer toward the photon orbit reflects the divergent relative Lorentz factor as circular timelike trajectories approach the null limit.

The right panel of Fig.~\ref{figDF} shows that $r_{\rm Rev}$ varies nonmonotonically with BH spin for the Hernquist spike. It crosses the prograde ISCO near $a/M\simeq0.32$, allowing the energy-gain region to overlap with stable circular orbits. We also consider the H\'enon isochrone model, a collisionless equilibrium with an analytic distribution function and adiabatic invariants~\cite{Henon:1959b,Henon:1960}. The energy-gain region outside the prograde ISCO persists for spikes grown from this model, showing that the effect is not specific to the Hernquist halo. The Supplemental Material derives a sufficient bound on occupation perturbations that preserve net energy gain and verifies its persistence across the scattering cutoffs tested. These results establish that negative dynamical friction does not require a finely tuned phase-space distribution.

\textit{Impact on quasi-circular inspirals}--- We quantify dynamical friction through inspiral dephasing and the conservative metric response through the circular-orbit redshift at fixed frequency. We evolve equatorial quasi-circular inspirals adiabatically \cite{Hughes:2005qb} using the orbital radius $r$ and $(2,2)$-mode carrier phase $\Phi_{22}$:
\begin{equation}
 \dot r=\frac{\dd r}{\dd  E} ( \dot E_{\rm GW} + \dot E_{\rm DF}),
 \ \ \ \ \dot \Phi_{22} = 2  \Omega ,
 \label{eq:adiabatic_evolution}
\end{equation}
where dots denote derivatives with respect to $t$. Both $\dot E_{\rm GW}$ and $\dot E_{\rm DF}$ are signed rates of change of the companion's specific orbital energy, including the perturbation parameters in Eq.~(\ref{eq:companion_motion}). We use $\dot E_{\rm GW}=-(\mathcal F_\infty+\mathcal F_{\rm H})/m$, with GW energy fluxes to infinity and into the horizon computed from the Kerr Teukolsky equation~\cite{Teukolsky:1973,BHPToolkit,TeukolskyToolkit}. The circular-geodesic quantities $\dd r/\dd E$ and $\Omega(r)$ are evaluated on vacuum Kerr. Equation~(\ref{eq:conserved_quantities_dot_dynamical_friction}) gives $\dot E_{\rm DF}$, with positive values denoting energy gain. The vacuum and ``2ndGSF'' evolutions set $\dot E_{\rm DF}=0$; the latter adds only a 3PN estimate of the orbit-averaged second-order dissipative energy rate to the leading-order vacuum model~\cite{Isoyama:2012Dephasing}. The orbit-averaged second-order dissipative contribution enters the $O(q^0)$ phase and is required at first post-adiabatic order, together with first-order self-force corrections~\cite{Hinderer:2008TwoTimescale}; the 3PN estimate thus serves as a benchmark for environmental energy transfer.

\begin{figure*}[!t]
	\centering
	\includegraphics[width=0.465\textwidth]{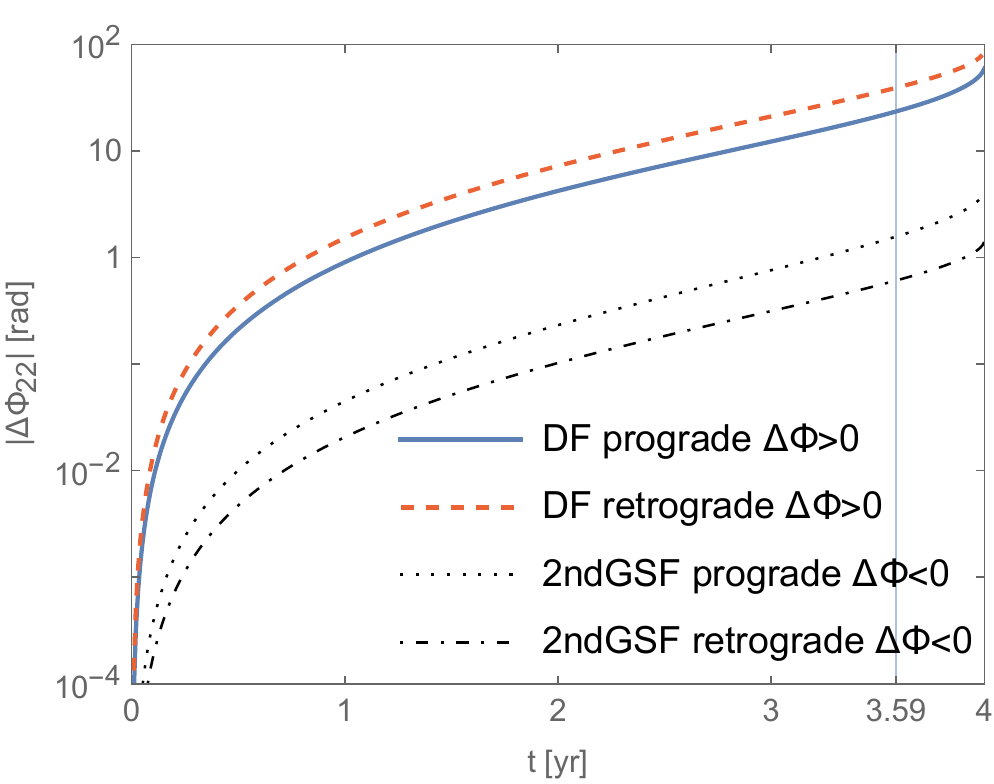}\hfill
	\includegraphics[width=0.49\textwidth]{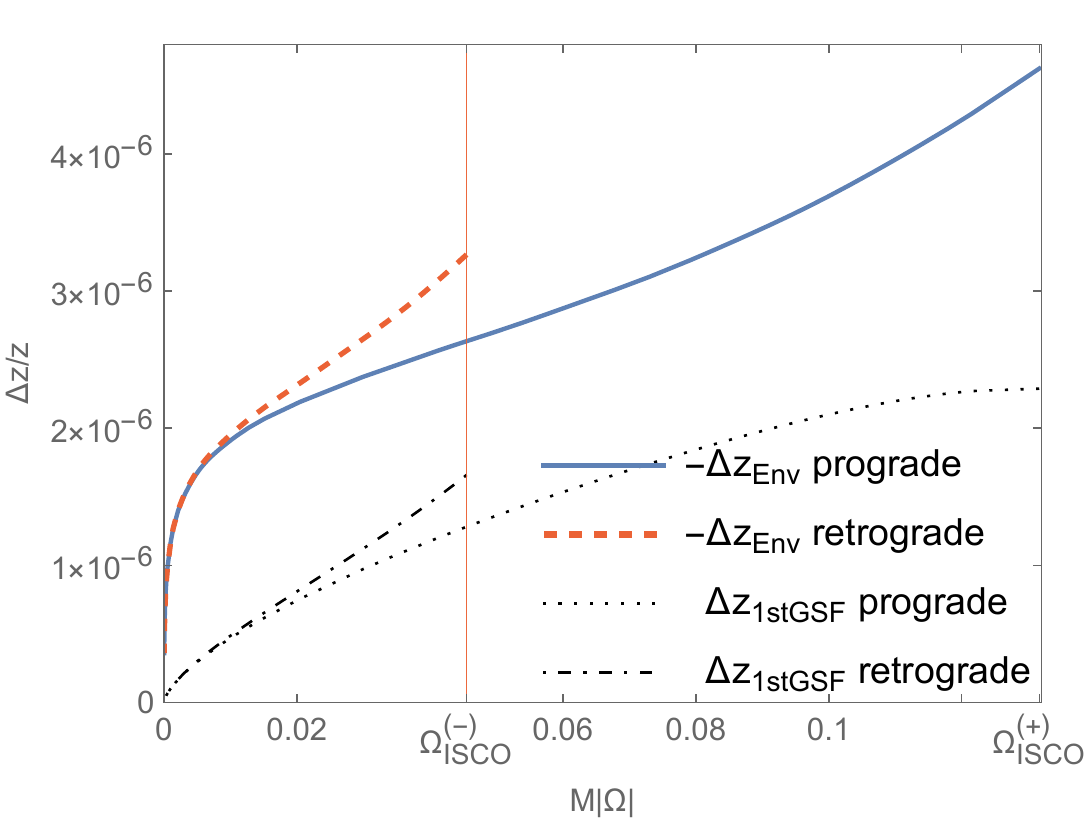}
	\caption{Inspiral dephasing and conservative redshift corrections for the parameters in Eq.~(\ref{eq:Milky Way-like}) and $q=10^{-5}$. Left: $|\Delta\Phi_{22}|$ relative to vacuum over four years for DF (blue solid: prograde; red dashed: retrograde) and the 3PN second-order dissipative self-force estimate (``2ndGSF''~\cite{Isoyama:2012Dephasing}; black dotted: prograde; black dot-dashed: retrograde). The vertical line at $t\simeq3.59\,{\rm yr}$ indicates the onset of prograde energy gain. Right: fixed-frequency redshift corrections, $-\Delta z_{\rm Env}/z_{\rm vac}$ and $\Delta z_{\rm 1stGSF}/z_{\rm vac}$ (8.5PN~\cite{Kavanagh:2016Redshift}), with curve styles given in the legend. Environmental corrections are plotted with reversed sign; the horizontal axis is $M|\Omega|$.}
	\label{figDiviation}
\end{figure*}

Figure~\ref{figDiviation} (left) compares four-year prograde and retrograde inspirals starting at $r/M\simeq7.85$ and $11.01$, respectively, and approaching the corresponding ISCOs. For each orientation, the corrected and vacuum evolutions start at the same orbital frequency and carrier phase. We define $\Delta\Phi_{22}^{X}(t)=\int_{t_i}^{t}[\omega_{22}^{X}(t')-\omega_{22}^{\rm vac}(t')]\,\dd t'$, where $\omega_{22}=2\Omega$ and $X=\mathrm{DF}$ or $\mathrm{2ndGSF}$ labels the correction. Dynamical friction produces phase differences of $\mathcal O(10^2)$ radians, compared with $\mathcal O(1)$ radians from the 3PN self-force estimate. For the prograde inspiral, this dephasing accumulates over both the energy-loss and energy-gain stages.

We assess the conservative metric response separately through the redshift at fixed orbital frequency. For equatorial circular geodesics of $g_{\mu\nu}^{\rm Env}=g_{\mu\nu}+\epsilon h_{\mu\nu}^{(1,0)}$, eliminating $r$ between $z_{\rm Env}(r)\equiv\dd\tau/\dd t$ and $\Omega_{\rm Env}(r)$ gives $z_{\rm Env}(\Omega)$. We obtain the first-order conservative self-force correction from the 8.5PN expansion of Ref.~\cite{Kavanagh:2016Redshift}. For each orbital orientation, we define
\begin{equation}
 \Delta z_X(\Omega)\equiv z_X(\Omega)-z_{\rm vac}(\Omega),
 \label{eq:redshift_correction}
\end{equation}
where $X=\mathrm{Env}$ or $\mathrm{1stGSF}$. Each comparison fixes $\Omega$ and the background Kerr parameters $M,a$, with time normalized at infinity. The corresponding orbital radii generally differ; the environmental and reference spacetimes need not have equal ADM charges. We normalize by $z_{\rm vac}(\Omega)$.

For the adopted spike model and mass ratio, the environmental and first-order conservative self-force corrections are comparable in magnitude but opposite in sign [Fig.~\ref{figDiviation} (right)], including at the initial prograde-inspiral frequency $M|\Omega|\simeq0.044$. The environmental redshift corrections for prograde and retrograde orbits nearly coincide at low frequencies but differ as the frequency approaches $|\Omega^{(-)}_{\rm ISCO}|$; the gravitational self-force corrections behave similarly.

{ \textit{Discussion}} ---
Negative dynamical friction transfers energy from DM particles to the secondary, partially offsetting GW losses and slowing the inspiral at a given orbital frequency. Within the local momentum-exchange model, the Supplemental Material establishes that energy gain survives a finite range of changes in phase-space occupation. Its occurrence for both cusped and cored halo progenitors further shows that the effect is not tied to one profile, although its strength and radial extent depend on the BH spin and the distribution. A Newtonian analogue is the acceleration of corotating eccentric perturbers near apocenter in rotating stellar disks~\cite{Bonetti:2020}. Here, energy gain occurs on stable circular orbits in a relativistic spike. This local momentum exchange has analogues in nonequilibrium negative drag~\cite{Pei:2025,Foffano:2012,Knippenberg:2025}and is distinct from wave amplification in rotating flows~\cite{Torres:2017} and BH superradiance~\cite{Cardoso:2011}. For the adopted spike and mass ratio, the fixed-frequency environmental redshift correction reaches the first-order conservative self-force scale even though the metric perturbation to Kerr is weak.

Self-consistent EMRI waveforms require the coupled metric and distribution responses $h_{\mu\nu}^{(1,1)}$ and $\mathcal F^{(1,1)}$, including their effects on radiation and orbital motion. The identification of wake-induced dynamical friction as part of the environmental gravitational self-force~\cite{Datta:2026DF} motivates relating our local scattering prescription to these retarded responses, with accretion treated consistently. Extending the calculation to eccentric and inclined orbits would test whether energy gain survives orbital averaging. Spinning secondaries would connect the framework to mass--spin coevolution in DM spikes~\cite{Karydas:2025bkj} and spin-dependent forces in gaseous EMRIs~\cite{Lui:2026SpinningSecondaries}.

\begin{acknowledgments}
This work makes use
of the \textit{Black Hole Perturbation Toolkit} and was supported in part by NSFC Grant No.12588101  and No.~12205104. This work is financially supported by the Ningbo Science and Technology Bureau under program grant No. 2024DJRC010009.
\end{acknowledgments}

%\bibliographystyle{apsrev4-1}
%\bibliography{DMBHrotating}

\vspace{10pt}
%%%%%%%%%%%%%%%%%%%%%%%
\section*{End Matter}
%%%%%%%%%%%%%%%%%%%%%%%
%
%%%%%%%%%%%%%%%%%%%%%%%

\subsection{Kinetic moments and dynamical friction}
\label{app:kinetic}
We work in Boyer--Lindquist coordinates $(t,r,\theta,\varphi)$ and define $\Sigma \equiv r^2+a^2\cos^2\theta$, $\Delta \equiv r^2-2Mr+a^2$. We introduce the mass integration variable $\mu'=\sqrt{-p_\alpha p^\alpha}$ and write $p^\alpha=\mu'v^\alpha$. For a fixed choice of the signs of the radial and polar velocities, the momentum volume element in Eq.~(\ref{eq:integral_energy_momentum}) can be expressed in terms of the conserved quantities as 
\begin{eqnarray}
\sqrt{-g} \ \dd ^4p &=& \sqrt{-g}\left|
 \frac{\partial(p^t,p^r,p^\theta,p^\varphi)}
      {\partial(\mathcal E,\mathcal{C},l_z,\mu')}
 \right| \dd \mathcal E \ \dd \mathcal{C} \ \dd l_z \ \dd \mu' \nonumber \\
 &=&\frac{\mu'^3\dd \mathcal E \ \dd \mathcal{C} \ \dd l_z \ \dd \mu' }{2 |v^r| |v^\theta| \ \Sigma^2  \sin\theta },
 \label{eq:Jacobian_conserved_quantities}
\end{eqnarray}
which reproduces the corresponding expression in Ref.~\cite{Ferrer:2017xwm}. For particles of fixed rest mass $\mu$, we write $\mathcal{F}^{(1,0)}=\mu'^{-4}\delta(\mu'-\mu)f^{(1,0)}$. This corresponds to the mass-shell number distribution $\mu^{-3}f^{(1,0)}$ used in Ref.~\cite{Vicente:2025gsg}, since $\sqrt{-g}\,\dd^4p=\mu'\,\dd\mu'\,\dd P$, with $\dd P=\dd^3p^{\bar i}/p^{\bar0}$ in a local orthonormal frame. We use the Carter-constant convention $\mathcal{C}=v_\theta^2+\frac{l_z^2}{\sin^2\theta}+a^2(1-\mathcal E^2)\cos^2\theta$. Thus $\mathcal{C}$ reduces to the square of the total specific angular momentum in the Schwarzschild limit. The companion constant $C$ follows the same convention, with $(v^\mu,\mathcal E,l_z)$ replaced by $(u^\mu,E,L_z)$. The velocity components $v^r$ and $v^\theta$ are given by
\begin{equation}
    v^r = \pm \frac{r^2 \sqrt{\mathcal{R}(r)}}{\Sigma},
    \qquad
    v^\theta = \pm\frac{\sqrt{\Theta(\theta)}}{\Sigma},
    \label{eq:ur_utheta_geodesic}
\end{equation}
with the radial and polar potentials
\begin{eqnarray}
	\mathcal{R}(r) &=& \mathcal{E}^2 \left(\frac{a^2 (2 M+r)}{r^3}+1\right) \nonumber \\
	&&-\frac{4 a M \mathcal{E} l_z}{r^3}+\frac{a^2 l_z^2}{r^4}-\frac{ \Delta\left(\mathcal{C}+r^2\right)}{r^4}
	\label{eq:radial_potential}, \\
	\Theta(\theta) &=& \mathcal{C}+(\mathcal{E}^2-1)a^2 \cos^2\theta - \frac{l_z^2}{\sin^{2}\theta },
	\label{eq:angular_potential}
\end{eqnarray}
respectively. The allowed particle orbits satisfy $\mathcal{R}(r)\geq0$ and $\Theta(\theta)\geq0$. Summing over both signs of $v^r$ and $v^\theta$ in Eq.~(\ref{eq:ur_utheta_geodesic}) cancels all off-diagonal components of the energy-momentum tensor except $\mathcal{T}_{t\varphi}$ and $\mathcal{T}_{\varphi t}$, and yields a factor of $4$ for the surviving components.

After integrating over $\mu'$, we obtain the energy-momentum tensor and the mass current (with the overall factor $\epsilon$ suppressed) as
\begin{eqnarray}
	\mathcal{T}_{\mu\nu} &=& \sum_{\pm v_r, \ \pm v_\theta} \int_{\mathcal{B}(r,\theta)}\dd \mathcal E \ \dd \mathcal{C} \ \dd l_z    \frac{v_\mu v_\nu \ \mu f^{(1,0)}}{2|v^r| |v^\theta| \ \Sigma^2  \sin\theta },  \nonumber \\
	&& \label{eq:expression_energy_momentum_tensor} \\
	J_{\mu} &\equiv& \mu\int_{\mathcal{B}(x)} p_\mu \mathcal{F}^{(1,0)} \sqrt{-g} \ \dd ^4p \nonumber \\
	&=& \sum_{\pm v_r, \ \pm v_\theta}\int_{\mathcal{B}(r,\theta)}\dd \mathcal E \ \dd \mathcal{C} \ \dd l_z    \frac{v_\mu \ \mu f^{(1,0)}}{2|v^r| |v^\theta| \ \Sigma^2  \sin\theta }, \nonumber \\
	&& \label{eq:expression_mass_current}
\end{eqnarray}
respectively. The reduced mass distributions used in Refs.~\cite{Sadeghian:2013laa,Ferrer:2017xwm} correspond to $\epsilon\mu f^{(1,0)}$ in our notation. The rest-mass density is defined by $\rho_M=\sqrt{-J_\mu J^\mu}$. Here the integration domain $\mathcal{B}(r,\theta)$, for noncaptured bound particles passing through $(r,\theta)$, is determined by the constraints imposed by the radial and polar potentials in Eqs.~(\ref{eq:radial_potential}) and~(\ref{eq:angular_potential}). The two largest real roots of $\mathcal{R}(r)$ are required to lie outside the event horizon, while circular orbits correspond to a degenerate root. The polar potential $\Theta(\theta)$ must possess at least one real root between the poles $\theta=0,\pi$. We use a numerical orbit-classification routine to determine the allowed region in conserved-quantity space; see Ref.~\cite{Ferrer:2017xwm} for a detailed discussion.

We use the ZAMO tetrad $(\bm{e}_{\hat{t}},\bm{e}_{\hat{r}},\bm{e}_{\hat{\theta}},\bm{e}_{\hat{\varphi}})$ to express the locally measured components of the energy-momentum tensor~\cite{Bardeen:1972}
\begin{eqnarray}
	&& \rho_{E} \equiv \mathcal{T}_{\hat{t}\hat{t}}, \ \  P_{\hat{r}} \equiv \mathcal{T}_{\hat{r}\hat{r}}, \ \  P_{\hat{\theta}} \equiv \mathcal{T}_{\hat{\theta}\hat{\theta}}, \nonumber \\
	&& P_{\hat{\varphi}} \equiv \mathcal{T}_{\hat{\varphi}\hat{\varphi}}, \ \ q_{\hat{\varphi}} \equiv - \mathcal{T}_{\hat{t}\hat{\varphi}}.
	\label{eq:densities_ZAMO}
\end{eqnarray}
Here $q_{\hat{\varphi}}$ is the azimuthal energy flux measured by ZAMOs, equivalently the azimuthal momentum density in units with $c=1$. It characterizes the net azimuthal transport of particle energy relative to these observers and does not by itself imply dissipative heat transport.

The energy conditions follow directly from the nonnegative distribution $\mathcal F^{(1,0)}$ in Eq.~(\ref{eq:integral_energy_momentum}). Wherever matter is present, the leading kinetic tensor admits an orthonormal eigenframe with energy density $\rho_*$ and principal pressures $P_{i*}$, $i=1,2,3$. In this frame, the kinetic integral and the mass-shell relation $p_\mu p^\mu=-\mu^2$ give
\begin{equation}
 \begin{gathered}
 P_{i*}\geq0,\\
 \rho_*-\sum_{i=1}^3P_{i*}
 =\mu^2\int_{\mathcal B(x)}\mathcal F^{(1,0)}\sqrt{-g}\,\dd^4p\geq0.
 \end{gathered}
\end{equation}
These inequalities imply that the null, weak, strong, and dominant energy conditions are satisfied. The eigenframe generally differs from the ZAMO frame.

To evaluate the dynamical friction in Eq.~(\ref{eq:acceleration_dynamical_friction}), we use the accretion and scattering cross sections in the local frame defined by the companion's Marck tetrad $(\bm u,\bm e_{\bar i})$. The momentum volume element is accordingly transformed into local velocity coordinates as
\begin{eqnarray}
\sqrt{-g} \ \dd ^4p &\triangleq&  \frac{\mu'}{p^{\bar{0}}}\dd \mu' \ \dd ^3p^{\bar{i}} \nonumber \\
 &\triangleq& \frac{1}{\gamma_{\bar{v}}} \left|\frac{\partial(p^{\bar{1}},p^{\bar{2}},p^{\bar{3}})}{\partial(v^{\bar{1}},v^{\bar{2}},v^{\bar{3}})}\right| \dd \mu' \ \dd ^3v^{\bar{i}} \nonumber \\
 &\triangleq& \mu'^3 \gamma_{\bar{v}}^4 \ \dd \mu' \ \dd ^3v^{\bar{i}}.
 \label{eq:Jacobian_Marck_velocity}
\end{eqnarray}
The local particle number density is $\dd N/(\dd V\,\dd^3v^{\bar i})=\epsilon f^{(1,0)}\gamma_{\bar v}^5$, obtained by integrating $p^{\bar0}\epsilon\mathcal F^{(1,0)}\sqrt{-g}\,\dd^4p$ over $\mu'$.
Using the companion's four-acceleration $\nabla_{\bm{u}}\bm{u} = \frac{1}{m} \left[\nabla_{\bm{u}}(m\bm{u})-\frac{\dd m}{\dd \tau}\bm{u}\right]$, we write the momentum-exchange contribution in Eq.~(\ref{eq:acceleration_dynamical_friction}) as 
\begin{eqnarray}
	\epsilon \ q \ \mathcal{A}^{\mu}_{\rm DF} &\triangleq& \frac{1}{m} \int_{\mathcal{B}(x)}  \left[\bar{v} \ \sigma_{\textrm{acc}} + \bar{v} \ \sigma_{\textrm{scat}} \right] \  v^{\bar i}e^\mu_{\bar i} \nonumber \\
	&& \times \epsilon \ \mu \ f^{(1,0)} \gamma_{\bar{v}}^6 \ \dd ^3v^{\bar{i}},
\end{eqnarray}
where the accretion and scattering cross sections $\sigma_{\textrm{acc}}(\bar{v})$ and $\sigma_{\textrm{scat}}(\bar{v})$ for environmental particles incident on a Schwarzschild BH are given explicitly in Ref.~\cite{Vicente:2025gsg}.

\subsection{Adiabatic growth of a rotating BH}
We assume that a BH grows adiabatically within an initial halo over a timescale much longer than the typical orbital period of the environmental particles. Under this adiabatic approximation, the particle actions in the initial model \cite{Sadeghian:2013laa,Gondolo:1999ef}
\begin{eqnarray}
    I^{(0)}_{r} &  =& \oint \sqrt{2\Psi^{(H)}-2\hat{\mathcal{E}}-\hat{l}^{2}/r^{2}} \ \dd r,  \nonumber \\
    I^{(0)}_{\theta} & =& 2\pi (\hat{l}-|l_{z}|),  \nonumber \\
    I^{(0)}_{\varphi} & =& 2\pi l_{z},
    \label{eq:Hernquist_adiabatic_invariants}
\end{eqnarray}
are adiabatic invariants: they remain constant as the BH grows. Here $\Psi^{(H)}$ is the relative potential of the Hernquist model;  the construction carries over to other ergodic collisionless models, such as the H\'enon isochrone, once the corresponding relative potential is specified. The particle actions after the adiabatic growth of a Kerr BH are given by
\begin{eqnarray}
    I^{(1)}_{r} &\equiv \oint v_{r} \dd r&=\oint \frac{r^2}{\Delta} \sqrt{\mathcal{R}} \dd r,  \nonumber \\
    I^{(1)}_{\theta} &\equiv \oint v_{\theta} \dd \theta&=\oint \sqrt{\Theta} \dd \theta,  \nonumber \\
    I^{(1)}_{\varphi} &\equiv \oint v_{\varphi} \dd \varphi&=2\pi l_{z},  
    \label{eq:Kerr_adiabatic_invariants}
\end{eqnarray}
with $v_{\varphi}=l_z$ and $v_{r}, \ v_{\theta}$ obtained from Eq.~(\ref{eq:ur_utheta_geodesic}). The conservation of adiabatic invariants gives
\begin{equation}
    I^{(0)}_{r}=I^{(1)}_{r}, \quad I^{(0)}_{\theta}=I^{(1)}_{\theta}, \quad I^{(0)}_{\varphi}=I^{(1)}_{\varphi},
    \label{eq:conservation_adiabatic_invariants}
\end{equation}
which implicitly determines $\hat{\mathcal{E}}(\mathcal{E},\mathcal{C},l_z)$. Conservation of particle number within a phase-space volume element gives rise to the adiabatic invariance of the distribution function \cite{Ferrer:2017xwm}
\begin{equation}
	f^{(H)}[\hat{\mathcal{E}}(\mathcal{E},\mathcal{C},l_z)] = f^{(1,0)}(\mathcal{E},\mathcal{C},l_z),
	\label{eq:adiabatic_invariance_distribution_function}
\end{equation}
where the Hernquist distribution function is obtained by Abel inversion. In Ref.~\cite{Hernquist:1990be}, $f(E)$ denotes the mass density in phase space. In our notation, the distribution is
\begin{eqnarray}
    \epsilon \ \mu \ f^{(H)}(\hat{\mathcal{E}}) = \frac{3 M_0^4  \tan ^{-1}\left(\sqrt{\frac{r_0\hat{\mathcal{E}}}{M_0-r_0 \hat{\mathcal{E}}}}\right)}{8 \sqrt{2} \pi ^3 r_0^{3/2} M_0^2 (M_0-r_0 \hat{\mathcal{E}})^{5/2}}& \nonumber \\
    +\frac{\sqrt{\hat{\mathcal{E}}} (2 r_0 \hat{\mathcal{E}}-M_0) \left(-8 r_0^2 \hat{\mathcal{E}}^2+8 r_0 M_0 \hat{\mathcal{E}}+3 M_0^2\right)}{8 \sqrt{2} \pi ^3 r_0 M_0^2 (r_0 \hat{\mathcal{E}}-M_0)^2} & .
    \label{eq:Hernquist_f}
\end{eqnarray}
The halo parameters in Eq.~(\ref{eq:Milky Way-like}) fix the physical normalization of $\epsilon\,\mu f^{(H)}$; $\epsilon$ is retained as a bookkeeping parameter.
% Subsection spacing separates the preceding display from the heading.

\subsection{Spectral construction of the metric response}
\label{app:metric}

We employ a stationary axisymmetric ansatz with circularity
\begin{equation}
 \begin{aligned}
 \dd   s^2={}&
 -e^{2F_0}N\,\dd t^2
 +e^{2F_1}\left(\frac{\dd \tilde{r}^2}{N}
                       +\tilde{r}^2\dd \theta^2\right)\\
 &+e^{2F_2}\tilde{r}^2\sin^2\theta
 \left(\dd \varphi-\frac{W}{\tilde{r}^2}\dd t\right)^2,
 \end{aligned}
 \label{eq:app_metric_ansatz}
\end{equation}
with $N=1-\tilde{r}_{\rm H}/\tilde{r}$. We hold the event horizon parameter at its background Kerr value $\tilde{r}_{\rm H} = 2\sqrt{M^2-a^2}$, while allowing the metric perturbations to modify the horizon area and intrinsic geometry. The metric functions $F_0,F_1,F_2,W$ depend on $(\tilde{r},\theta)$.
This ansatz is widely used in the numerical construction of rotating BH solutions.
The Kerr background functions $(F^{\rm Kerr}_0,F^{\rm Kerr}_1,F^{\rm Kerr}_2,W^{\rm Kerr})$ follow from rewriting the Boyer--Lindquist metric with $r=\tilde{r}+M-\sqrt{M^2-a^2}$. 

To obtain the linearized Einstein equations~(\ref{eq:Einstein_10_environment}), we expand
\begin{equation}
 F_i=F_i^{\rm Kerr}+\epsilon\,\delta F_i
,\qquad
 W=W^{\rm Kerr}+\epsilon\,\delta W\ccr{.}
 \label{eq:app_metric_expansion}
\end{equation}
The coefficients of $\epsilon$ in the expanded metric components define $h_{\mu\nu}^{(1,0)}$ in the main text.
The functions ($\delta F_0,\delta F_1,\delta F_2,\delta W$) are determined by the energy-momentum tensor in Eq.~(\ref{eq:expression_energy_momentum_tensor}) through the linear equations in Eq.~(\ref{eq:Einstein_10_environment}). We discretize these equations using a spectral method~\cite{Fernandes:2022gde} and solve the resulting system by Newton--Raphson iteration, starting from the Kerr seed $\delta F_i=\delta W=0$. To map the radial domain $\tilde{r} \in [\tilde{r}_{\rm H},\infty)$ to $x \in [-1,1]$, we use the compactifying transformation \cite{Guo:2023ScalarizedKerrNewman}
\begin{equation}
    x\equiv\frac{\sqrt{\tilde{r}^{2}-\tilde{r}_{\rm H}^{2}}-\tilde{r}_{\rm H}}{\sqrt{\tilde{r}^{2}-\tilde{r}_{\rm H}^{2}}+\tilde{r}_{\rm H}}.
    \label{eq:app_transformation_r}
\end{equation}
The boundary conditions at the event horizon are
\begin{equation}
 \left.
 (\partial_x\delta F_i)
 \right|_{x=-1}
 =(0,0,0), \ \ \delta W|_{x=-1}=0.
 \label{eq:app_inner_BC}
\end{equation}
We impose $\delta\Omega_H=0$ to select a stationary solution family with fixed horizon angular velocity relative to the nonrotating frame at infinity, while allowing the horizon angular momentum to respond to the environment. Asymptotic flatness requires the following boundary conditions at infinity
\begin{equation}
(\delta F_i,\delta W)|_{x=1}\rightarrow(0,0,0,0).
\label{eq:app_infinity_BC}
\end{equation}
Axis regularity and equatorial reflection symmetry require $\partial_\theta\delta F_i=\partial_\theta\delta W=0$ at $\theta=0,\ \pi/2$, together with $\delta F_1|_{\theta=0} = \delta F_2|_{\theta=0}$ to remove conical singularities on the symmetry axis.

% Supplemental Material continues below End Matter in a single column.
\onecolumngrid
\setcounter{equation}{0}
\setcounter{figure}{0}
\setcounter{table}{0}
\setcounter{section}{0}
\renewcommand{\theequation}{S\arabic{equation}}
\renewcommand{\thefigure}{S\arabic{figure}}
\renewcommand{\thetable}{S\arabic{table}}
% Give reset counters distinct hyperlink destinations.
\providecommand{\theHequation}{}
\providecommand{\theHfigure}{}
\providecommand{\theHtable}{}
\providecommand{\theHsection}{}
\renewcommand{\theHequation}{supp.\arabic{equation}}
\renewcommand{\theHfigure}{supp.\arabic{figure}}
\renewcommand{\theHtable}{supp.\arabic{table}}
\renewcommand{\theHsection}{supp.\arabic{section}}
\section*{Supplemental Material}
% !TeX root = DMBHrotating.tex
% Included by DMBHrotating.tex; main.tex remains the standalone wrapper.
% Numerical diagnostics independently recomputed on 2026-11-05.
% Reproducible code, raw refinements, and validation records:
% ../numerical-validation/constant, /kernel, and /hernquist.
% This independent recomputation does not constitute a run of the authors'
% original production code, which was not available for this check.
% Requires amsmath, amssymb, bm, graphicx, and xcolor.
% Bibliography keys already occur in the main manuscript.

\providecommand{\ccr}[1]{{\color{red}#1}}
% Red text marks the editorial revision of 2026-11-07.

In this Supplemental Material, we explain how dark matter particles overtaking the companion in azimuth can supply orbital energy. Three ingredients enter the local momentum-exchange prescription: the allowed Kerr orbital domain, its phase-space occupation, and the velocity dependence of the interaction kernel. Kerr's binding and capture conditions allow different velocity ranges for particles moving faster and slower than the companion in azimuth. To isolate the roles of orbital support and interaction weighting, Sec. I considers a constant phase-space distribution. In this example, enhanced scattering at small relative speeds makes the energy-supplying contribution exceed the losses. Section II derives a sufficient bound on occupation changes that preserve energy gain, while Sec. III tests halo models and scattering cutoffs. These results show that negative dynamical friction requires neither a finely tuned distribution nor a specific halo profile, although its strength and radial extent remain model-dependent.

\section{Energy-supply mechanism}
\label{sec:ndfr-mechanism}

Consider a nonspinning companion of mass \(m\) on a stable prograde equatorial circular orbit around a Kerr black hole of mass \(M\) and spin parameter \(a>0\). Write its four-velocity as \(\bm u=u^t(\bm\partial_t+\Omega\bm\partial_\varphi)\), where \(\Omega\equiv\mathrm d\varphi/\mathrm dt=u^\varphi/u^t=\sqrt M/(r^{3/2}+a\sqrt M)\) is the orbital angular velocity with respect to Boyer--Lindquist time \(t\). Denote its specific energy and angular momentum by \(E=-u_t\) and \(L_z=u_\varphi\). The local azimuthal basis vector may be chosen as
\begin{equation}
 \begin{aligned}
 \bm e_{\bar3}&=\frac{\bm\partial_\varphi+L_z\bm u}{D},
 \qquad -(e_{\bar3})_t=\Omega D,\\
 D&=\sqrt{g_{\varphi\varphi}+L_z^{\,2}}.
 \end{aligned}
 \label{eq:ndfr-tetrad}
\end{equation}
Let \(w=v^{\bar3}\), \(\bm v_\perp=(v^{\bar1},v^{\bar2})\), \(\bar v^2=v_\perp^2+w^2\), and \(\gamma_{\bar v}=(1-\bar v^2)^{-1/2}\). We use \(G=c=1\).

At the companion's position, the indicator \(\chi_{\mathcal B}\) selects local velocities corresponding to bound, noncaptured Kerr geodesics, including inclined orbits crossing the equatorial plane~\cite{Ferrer:2017xwm}. Although the prograde marginally bound orbit sets the population's inner edge, membership in \(\mathcal B\) requires the full energy- and angular-momentum-dependent noncapture condition.

Writing \(f\equiv f^{(1,0)}\), the local momentum-exchange prescription~\cite{Vicente:2025gsg} gives
\begin{align}
 \dot E_{\rm DF}[f]
 &=\mathcal A\int \chi_{\mathcal B}(v^{\bar i})
 W(\bar v)\,w\,f(v^{\bar i})\,\mathrm d^3v^{\bar i},
 \label{eq:ndfr-rate}
\end{align}
\begin{equation}
 \begin{aligned}
 W(\bar v)&=\bar v\gamma_{\bar v}^6
 \bigl[\sigma_{\rm acc}(\bar v)+\sigma_{\rm scat}(\bar v)\bigr]\geq0,
 \\
 \mathcal A&=\frac{\epsilon\mu\Omega D}{m u^t}>0 .
 \end{aligned}
 \label{eq:ndfr-kernel}
\end{equation}
Here \(\mu\) is the particle rest mass, \(\epsilon\) is the environmental expansion parameter, and the dot denotes differentiation with respect to Boyer--Lindquist time. Both scattering and accretion are included, as in the main text. Since \(\mathcal A>0\), positive azimuthal momentum transfer increases the companion's specific orbital energy. For a fixed background, orbit, and impact-parameter prescription, \(\chi_{\mathcal B}\) and \(W\) are independent of the occupation \(f\).

Extend the occupied distribution by zero outside \(\mathcal B\), defining \(F(v^{\bar i})=\chi_{\mathcal B}(v^{\bar i})f(v^{\bar i})\). Pairing opposite azimuthal velocities yields
\begin{equation}
 \begin{aligned}
 \dot E_{\rm DF}[f]
 &=\mathcal A\int_{w>0}W(\bar v)w\,
 \bigl[F(\bm v_\perp,w)\\
 &\qquad-F(\bm v_\perp,-w)\bigr]
 \,\mathrm d^2v_\perp\,\mathrm dw .
 \end{aligned}
 \label{eq:ndfr-pair}
\end{equation}
For each pair of opposite azimuthal velocities, the interaction kernel is identical. Their net contribution is therefore proportional to the difference in the occupied distribution \(F\) between the two velocities.

We first take \(f=f_0>0\) throughout \(\mathcal B\). This makes the invariant phase-space occupation constant. The local number distribution per unit velocity-space volume still depends on speed, since \(\mathrm dN/(\mathrm dV\,\mathrm d^3v^{\bar i})=\epsilon f_0\gamma_{\bar v}^5\). For this model, Eq.~\eqref{eq:ndfr-pair} becomes
\begin{align}
 \dot E_{\rm DF}[f_0]
 &=\mathcal A f_0\int_{w>0}W(\bar v)w\,\Delta\chi\,
 \mathrm d^2v_\perp\,\mathrm dw ,
 \label{eq:ndfr-constant}\\
 \Delta\chi
 &\equiv\chi_{\mathcal B}(\bm v_\perp,w)
       -\chi_{\mathcal B}(\bm v_\perp,-w).
 \nonumber
\end{align}
Here \(\Delta\chi\in\{-1,0,1\}\), and pairs for which both velocities are allowed cancel exactly. The regions with \(\Delta\chi=+1\) and \(-1\) supply and remove energy, respectively; net gain requires a positive \(Ww\)-weighted integral, not a pointwise positive \(\Delta\chi\).

\begin{figure*}[tp]
 \centering
 \includegraphics[width=\textwidth]{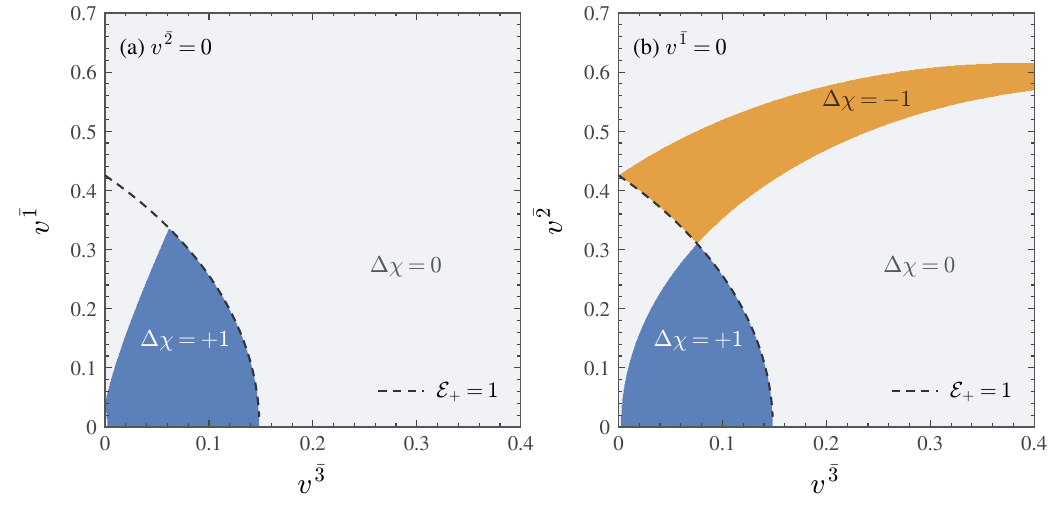}
 \caption{Paired orbital-support difference \(\Delta\chi\) at \(a/M=0.65\) and \(r/M=4\). Panels (a) and (b) fix \(v^{\bar2}=0\) and \(v^{\bar1}=0\), respectively, and show the positive half-space \(w=v^{\bar3}>0\). All velocity components are measured in the companion's local orthonormal frame, as in the main text. Blue, orange, and gray regions have \(\Delta\chi=+1,-1,0\), as marked inside the panels. The dashed curve is the positive member's bound-energy boundary, \(\mathcal E_+=1\). These slices illustrate the allowed orbital domain; energy transfer is evaluated over the complete three-dimensional velocity domain.}
 \label{fig:ndfr-support}
\end{figure*}

The competing restrictions can be understood from the particle's conserved specific energy \(\mathcal E\) and axial angular momentum \(l_z\), using the notation of the main text:
\begin{equation}
 \begin{aligned}
 \mathcal E_\pm&=\gamma_{\bar v}(E\pm\Omega Dw),\\
 l_{z,\pm}&=\gamma_{\bar v}(L_z\pm Dw).
 \end{aligned}
 \label{eq:ndfr-particle-constants}
\end{equation}
The paired velocities have different conserved energies and angular momenta. The positive member can remain bound and noncaptured while its negative counterpart plunges, giving \(\Delta\chi=+1\). Conversely, since \(\mathcal E_+>\mathcal E_-\), the positive member can exceed the bound-energy limit \(\mathcal E=1\) while its negative counterpart remains allowed, giving \(\Delta\chi=-1\). Figure~\ref{fig:ndfr-support} illustrates both possibilities. The relevant bias is measured relative to the companion, rather than by the sign of \(l_z\) alone.

\begin{samepage}
To quantify the balance, let \(\mathcal B_>=\mathcal B\cap\{w>0\}\) and \(\mathcal B_<=\mathcal B\cap\{w<0\}\), and define the supply-to-loss ratio over these full domains:

\begin{equation}
 R_K[f]=
 \frac{\displaystyle\int_{\mathcal B_>}K(\bar v)\,w f\,\mathrm d^3v^{\bar i}}
 {\displaystyle\int_{\mathcal B_<}K(\bar v)\,|w|f\,\mathrm d^3v^{\bar i}}.
 \label{eq:ndfr-halo-ratio}
\end{equation}
\end{samepage}
Here \(K=W\) gives the physical ratio, with \(R_W>1\) denoting net gain; \(K=1\) removes the velocity-dependent kernel solely as a diagnostic. Both ratios are independent of the overall normalization of \(f\).

For \(a/M=0.65\), \(q=m/M=10^{-5}\), \(r=4M\), and \(b_{\max}=r(q/3)^{1/3}\), constant occupation gives \(R_1[f_0]\simeq0.611\) but \(R_W[f_0]\simeq2.218\). The physical kernel thus reverses the balance from net loss to net gain. In this benchmark, the unpaired supply region preferentially samples lower particle speeds relative to the companion, whose contribution is enhanced by the scattering cross section. Energy gain reflects asymmetric orbital support combined with velocity-dependent interactions, rather than the size of the supply region alone. For the Hernquist spike at the same parameters, scattering supplies net orbital energy while accretion removes it; scattering dominates, as quantified in Sec.~\ref{sec:ndfr-halo}.

\section{Robustness under occupation perturbations}
\label{sec:ndfr-robustness}

For the constant-occupation benchmark, define the nonnegative supply and loss contributions and their normalized difference:

\begin{equation}
 \begin{aligned}
 P_0&=f_0\int_{\mathcal B_>}Ww\,\mathrm d^3v^{\bar i},\\
 N_0&=f_0\int_{\mathcal B_<}W|w|\,\mathrm d^3v^{\bar i},\\
 \mathcal D_0&=\frac{P_0-N_0}{P_0+N_0}.
 \end{aligned}
 \label{eq:ndfr-margin}
\end{equation}
The net rate is \(\mathcal A(P_0-N_0)\). Representative values are listed in Table~\ref{tab:ndfr-radii}. Energy gain extends from the prograde ISCO, \(r_{\rm ISCO}\simeq3.616M\), to \(r_{\rm Rev}\simeq4.93M\). These radii characterize the constant-distribution benchmark and need not coincide with those of a particular halo-derived spike.

\begin{table}[!htbp]
 \centering
 \caption{Normalized difference between energy supply and loss for a constant distribution on bound, noncaptured Kerr orbits. Positive values denote net energy gain.}
 \label{tab:ndfr-radii}
 \begingroup
 \setlength{\tabcolsep}{12pt}
 \renewcommand{\arraystretch}{1.15}
 \begin{tabular}{cc}
 \hline\hline
 \(r/M\) & \(\mathcal D_0\)\\
 \hline
 3.7 & \(+0.5145\)\\
 4.0 & \(+0.3785\)\\
 4.5 & \(+0.1663\)\\
 4.9 & \(+0.0119\)\\
 5.0 & \(-0.0244\)\\
 \hline\hline
 \end{tabular}
 \endgroup
\end{table}

The positive margin \(\mathcal D_0\) provides a sufficient condition for energy gain to persist under changes in the distribution. Keep the background, companion orbit, allowed domain \(\mathcal B\), and interaction kernel \(W\) fixed, and consider
\begin{equation}
 f=f_0(1+h),\qquad |h|\leq\eta<1
 \quad\hbox{on }\mathcal B .
 \label{eq:ndfr-occupation-perturbation}
\end{equation}
The perturbed distribution must remain physically admissible; for a stationary collisionless population, \(h\) may be taken to depend on the particle's conserved quantities. Nonnegativity of the weights in \(P_0\) and \(N_0\) implies
\begin{equation}
 \begin{aligned}
 \dot E_{\rm DF}[f]
 &\geq\mathcal A\big[(1-\eta)P_0-(1+\eta)N_0\big]\\
 &=\mathcal A(P_0+N_0)(\mathcal D_0-\eta)>0,
 \qquad \eta<\mathcal D_0 .
 \end{aligned}
 \label{eq:ndfr-robustness-bound}
\end{equation}
Thus a strictly positive margin establishes an open neighborhood of energy-gaining occupations, rather than an isolated distribution. The values in Table~\ref{tab:ndfr-radii} give sufficient bounds of approximately \(38\%\) at \(r=4M\) and \(17\%\) at \(r=4.5M\) on the pointwise relative occupation change. These percentages refer to \(f\), at fixed orbital support and interaction prescription, and are not bounds on variations of halo parameters. The halo-derived distributions below provide complementary tests beyond this neighborhood.

\section{Halo models and numerical checks}
\label{sec:ndfr-halo}

A nonconstant \(f\) changes the occupation of the allowed orbits and can prevent cancellation even when both members of a velocity pair are allowed. Adiabatic growth fixes the final occupation through \(f_{\rm final}(\bm I)=f_{\rm initial}(\bm I)\) for trajectories surviving capture, with \(\bm I=(I_r,I_\theta,I_\varphi)\)~\cite{Ferrer:2017xwm}. We denote by \(f_{\rm H}\) the spike grown from the main-text Hernquist halo, with \(M_0=387500M\) and \(r_0=227775M_0\).

\begin{table}[!htbp]
 \centering
 \caption{Effect of occupation and interaction weighting at \(a/M=0.65\), \(r=4M\), \(q=10^{-5}\), and \(b_{\max}=r(q/3)^{1/3}\). Ratios use the full domains \(\mathcal B_>\) and \(\mathcal B_<\); only \(R_W\) determines the physical sign.}
 \label{tab:ndfr-halo-weighting}
 \begingroup
 \setlength{\tabcolsep}{8pt}
 \renewcommand{\arraystretch}{1.15}
 \begin{tabular}{lcc}
 \hline\hline
 Distribution & \(R_1[f]\) & \(R_W[f]\)\\
 \hline
 Constant \(f_0\) & \(0.611\) & \(2.218\)\\
 Hernquist \(f_{\rm H}\) & \(0.476\) & \(3.131\)\\
 \hline\hline
 \end{tabular}
 \endgroup
\end{table}

Table~\ref{tab:ndfr-halo-weighting} shows that the Hernquist occupation lowers the constant-kernel ratio but increases the physical supply-to-loss ratio by about \(41\%\). The same kernel is sampled differently by the two distributions: occupation and velocity weighting together enhance the relative supply, without changing the allowed orbital domain. This comparison concerns the balance of supply and loss, not the absolute rates for differently normalized distributions.

The separate scattering and accretion terms clarify which interaction supplies energy. For the Hernquist spike, replacing \(K\) by \(\bar v\gamma_{\bar v}^6\sigma_{\rm scat}\) gives a supply-to-loss ratio of \(3.402\), while \(K=\bar v\gamma_{\bar v}^6\sigma_{\rm acc}\) gives \(0.436\). Scattering thus supplies energy, whereas accretion by the companion removes it. The net scattering contribution is approximately \(42.4\) times the magnitude of the net accretion loss, so scattering drives the positive total rate in this example.

The main-text calculations also find energy gain on stable prograde circular orbits for spikes grown from H\'enon isochrone halos. The Hernquist and isochrone progenitors have different inner density structures, a cusp and a core, respectively, and generate nonconstant occupations through the same adiabatic construction.

\medskip
\noindent\textit{Numerical prescription and cutoff check.}
The scattering kernel is evaluated from the exact Schwarzschild deflection angle, rather than a fit to the Coulomb logarithm:
\begin{equation}
 \begin{aligned}
 \sigma_{\rm acc}&=\pi b_{\rm cr}^{\,2},\\
 \sigma_{\rm scat}&=4\pi\int_{b_{\rm cr}}^{b_{\max}}
 b\cos^2\phi_\infty(b,\bar v)\,\mathrm db .
 \end{aligned}
 \label{eq:ndfr-exact-cross-sections}
\end{equation}
Here \(b_{\rm cr}(\bar v)\) is the companion's critical capture impact parameter and \(\phi_\infty\) is the angle accumulated from periapsis to infinity on a scattering orbit~\cite{Vicente:2025gsg}. The scattering integral is set to zero when \(b_{\max}\leq b_{\rm cr}\); the capture cross section is retained as written.

Varying \(b_{\max}\) from \(b_{\rm H}/2\) to \(2b_{\rm H}\), where \(b_{\rm H}=r(q/3)^{1/3}\), preserves net energy gain at both radii in Table~\ref{tab:ndfr-cutoff}. Before rounding, the largest estimated absolute quadrature error in \(\mathcal D_0\) is \(1.1\times10^{-5}\), well below the positive margins. We estimate this error as twice the larger absolute change across the last two quadrature refinements; it excludes uncertainty in the interaction model. For the Hernquist ratios in Table~\ref{tab:ndfr-halo-weighting}, refinement gives estimated absolute errors of \(10^{-4}\) in \(R_1\) and \(10^{-3}\) in \(R_W\).

\begin{table}[!htbp]
 \centering
 \caption{Sensitivity to the maximum scattering impact parameter for constant occupation at \(a/M=0.65\) and \(q=10^{-5}\), with \(b_{\rm H}=r(q/3)^{1/3}\). All six cases retain positive net energy transfer.}
 \label{tab:ndfr-cutoff}
 \begingroup
 \setlength{\tabcolsep}{10pt}
 \renewcommand{\arraystretch}{1.15}
 \begin{tabular}{ccc}
 \hline\hline
 \(r/M\) & \(b_{\max}/b_{\rm H}\) & \(\mathcal D_0\)\\
 \hline
 4.0 & \(1/2\) & \(0.3374\)\\
 4.0 & \(1\) & \(0.3785\)\\
 4.0 & \(2\) & \(0.4113\)\\
 4.5 & \(1/2\) & \(0.1238\)\\
 4.5 & \(1\) & \(0.1663\)\\
 4.5 & \(2\) & \(0.2001\)\\
 \hline\hline
 \end{tabular}
 \endgroup
\end{table}

% Keep the concluding paragraph after the final floating table.
\newpage

Together, these results show that asymmetric Kerr orbital support and velocity-dependent interaction weights can produce net energy gain even for constant phase-space occupation. The bound in Eq.~\eqref{eq:ndfr-robustness-bound} guarantees that this gain persists under finite, physically admissible occupation perturbations at fixed orbital support and interaction kernel, and the numerical tests confirm its persistence across the scattering cutoffs examined. Its realization in spikes grown from both cusped and cored halos further shows that the effect is not specific to a single halo profile. Its strength and radial extent nevertheless depend on the distribution and orbital parameters.

\end{document}